\documentclass[12pt]{article}
\usepackage[utf8]{inputenc}
\usepackage[T1]{fontenc}
\usepackage{lmodern}
\usepackage{amsmath,amssymb,amsthm}
\usepackage[authoryear,round]{natbib}
\usepackage[colorlinks=true,citecolor=blue,linkcolor=blue,urlcolor=blue]{hyperref}
\usepackage[margin=25mm]{geometry}
\usepackage{setspace}
\newtheorem{theorem}{Theorem}
\newtheorem{lemma}{Lemma}
\newtheorem{proposition}{Proposition}
\newtheorem{corollary}{Corollary}
\theoremstyle{definition}
\newtheorem{definition}{Definition}
\newtheorem{example}{Example}
\newtheorem{remark}{Remark}
\newcommand{\R}{\mathbb R}
\newcommand{\W}{\mathcal W}
\usepackage{xcolor}

\title{Strictly Speaking:\\ Log-Supermodularity, Single Crossing, and Transversality}
\author{Jaime Orrillo\thanks{Catholic University of Bras\'ilia. The author gratefully acknowledges the hospitality of IMPA (Instituto de Matem\'atica Pura e Aplicada), where part of this work was developed. The author has regularly participated in IMPA's summer program every year since 2000, except in 2020 due to the COVID-19 pandemic. E-mail: \texttt{orrillo@p.ucb.br}, \texttt{jjoc@impa.br}.}}
\date{November 2026}

\begin{document}
\maketitle

\begin{abstract}
Log-supermodularity of a kernel is commonly used to draw two conclusions about the integral of a single-crossing function: strict single crossing, and transversality, a strictly positive derivative at the crossing. We show that the first is characterized by strictly increasing likelihood ratios and the second by a strictly increasing score, and that the gap between them consists of crossings with zero derivative, which a derivative-free index condition captures. In the security-bid auctions of DeMarzo, Kremer and Skrzypacz (2005), the strict monotone likelihood ratio property therefore does not deliver the strict ranking of slopes of their Lemma~5, and we give a counterexample within their model. Their ranking of revenues across security designs nevertheless holds under that property, by strict single crossing in second-price auctions and by a comparison argument, under a regularity condition on the set of securities, in first-price auctions. The same argument extends to the common-value model of Yuan (2026), where it delivers the revenue ranking of informal and formal first-price auctions without the derivative conclusion.
\end{abstract}



\newpage

\begin{flushright}
\begin{minipage}{0.62\textwidth}
\small\itshape
Citius emergit veritas ex errore quam ex confusione.\\[0.5ex]
\upshape Truth will sooner come out from error than from confusion.\\[0.5ex]
--- Francis Bacon, \emph{Novum Organum} (1620), Book~II, Aphorism~XX
\end{minipage}
\end{flushright}
\vspace{2ex}

\section{Introduction}

Arguments in monotone comparative statics under uncertainty, auction theory, and security design often reduce to a sign question for an integral: whether an agent's choice is monotone in a signal \citep{Athey2002}, whether a bidder prefers one offer to another given his type \citep{MilgromWeber1982}, or whether one security is more sensitive than another to the type of the bidder \citep{DeMarzoKremerSkrzypacz2005}. In each case there is a function of a state that changes sign at most once, a positive kernel indexed by a parameter, and the object of interest is the integral of the function against the kernel, as a function of the parameter. The question is when the sign pattern of the integrand is transmitted to the integral, and in what form.

The classical answer is that log-supermodularity of the kernel preserves single crossing \citep{Karlin1968,Athey2002}. In applications, however, two different conclusions are drawn from it. The first is \emph{strict single crossing}: once the integral is non-negative it stays strictly positive. The second is \emph{transversality} at the crossing: where the integral vanishes, its derivative is strictly positive, so that zero is a regular value of the integral near the crossing.\footnote{~The term is used in the sense of differential topology: a zero of a function is transversal if the function crosses zero with non-zero derivative, so that zero is a regular value near it, as in the theory of regular economies of \citet{Debreu1970}. It is unrelated to the transversality condition of infinite-horizon optimization. Here the crossing is also oriented, with a positive derivative; Section~\ref{sec:index} describes the orientation without derivatives.} The second is what one needs to perturb an equilibrium, to apply an implicit-function argument, or to run local arguments of the type of \citet[Lemma~2]{MilgromWeber1982}. The two are frequently stated together, as consequences of the same hypothesis.

This paper separates them, characterizes each, and applies the distinction to security-bid auctions. The general results are the following.
\begin{itemize}
\item Strict single crossing of the integral, for every bounded, compactly supported, weakly single-crossing integrand that is non-zero on a set of positive measure, holds if and only if every likelihood ratio of the kernel is strictly increasing (Theorem~\ref{thm:A}). This is strict log-supermodularity in ratio form, and it involves no derivatives.
\item Transversality, for every such integrand, holds if and only if the score of the kernel, the derivative of its logarithm with respect to the parameter, is strictly increasing in the state (Theorem~\ref{thm:B}).
\item The score condition implies the likelihood ratio condition, but not conversely; the likelihood ratio condition yields only a weak ranking of derivatives at the crossing (Proposition~\ref{prop:hier} and Corollary~\ref{cor:weak}). A simple exponential kernel separates the two (Example~\ref{ex:sep}).
\item For continuous functions, strict single crossing is equivalent to a condition on local crossing indices: every zero is a change of sign from negative to positive (Theorem~\ref{thm:index}). No derivative is involved, and a crossing may have zero derivative. The argument is elementary, and can be read as the one-dimensional case of a degree argument.
\end{itemize}

In the auctions studied by \citet{DeMarzoKremerSkrzypacz2005}, henceforth DKS, bidders offer securities, and the ranking of revenues across security designs rests on how the expected payments of two securities cross as a function of the bidder's type. DKS call a security steeper than another when equality of expected payments implies a strict ranking of their slopes; their Lemma~5 states that this holds whenever the payoffs of the two securities cross once from below, and their Proposition~1 shows that steeper sets of securities yield higher revenues in first-price and second-price auctions. The conditional density of the payoff is assumed to satisfy the strict monotone likelihood ratio property, which the paper notes to be equivalent, under differentiability, to a strictly positive cross-partial derivative of the logarithm of the density. With strict inequalities, only one direction of this equivalence holds: this is the gap between likelihood ratios and scores described above.

We show that the gap matters for the lemma but not for the proposition. Example~\ref{ex:dks} gives a conditional density and two feasible securities that satisfy all the assumptions of DKS, including their normalization of the signal, whose payoffs cross once from below and whose expected payments are equal, with equal slopes, at some signal. The exact condition for the lemma is strict monotonicity of the score, which characterizes transversality (Theorem~\ref{thm:B}); the strictly positive cross-partial derivative of DKS is a familiar sufficient condition for it, but not a necessary one. In the terminology of \citet{Lakatos1976}, the example reveals a hidden lemma, strict monotonicity of the score. The revenue ranking, on the other hand, holds under the monotone likelihood ratio property as stated. For second-price auctions it follows from strict single crossing alone (Proposition~\ref{prop:spa}). For first-price auctions the argument in DKS is pointwise: at a type where the bidders' payoffs under the two designs coincide, the strict ranking of slopes ranks the derivatives of the payoffs. We replace it by a comparison argument for the differential inequality that the difference of payoffs satisfies, which needs only the weak ranking of slopes, at all payment levels, together with a regularity condition on the set of securities (Proposition~\ref{prop:fpa}). The ranking is strict unless slopes coincide on a set of types of positive measure.

The lemma has been used in later work. In the common-value model of \citet{Yuan2026}, Lemma~5 of that paper extends it to a general kernel and states both conclusions. Section~\ref{sec:yuan-use} explains that the main results there rely on strict single crossing, which holds under the stated hypotheses, and that the derivative conclusion is used in one place, the comparison of the informal auction with formal first-price auctions. The same comparison argument shows that this comparison holds under the likelihood ratio property, under a regularity condition on the set of securities (Proposition~\ref{prop:yuan}).

Preservation of sign patterns by totally positive kernels is the variation-diminishing property of \citet{Karlin1968}; its use in economics under uncertainty is systematized by \citet{Athey2002}. Monotone comparative statics rest on single crossing of incremental differences \citep{MilgromShannon1994} and its refinements \citep{QuahStrulovici2009}. \citet{KartikLeeRappoport2024} characterize the utility functions whose expectations have single-crossing differences across all distributions in a convex environment; the problem studied here is dual to theirs, since we fix the single-crossing integrand and characterize the kernel, and the derivative at the crossing plays no role in their analysis. Single crossing is not preserved by sums in general \citep{QuahStrulovici2012}, which is one reason why preservation under integration needs structure on the kernel. The ranking of auctions through the sensitivity of payments to the winner's type is the linkage principle of \citet{MilgromWeber1982}. Earlier comments on DKS address different issues: adverse selection when the investment cost is correlated with returns \citep{CheKim2010}, and risk aversion and weaker forms of positive dependence \citep{AHW2015}. As far as we know, the distinction studied here has not been raised.

We note the limits of the analysis. For first-price auctions we take the characterization of equilibrium in DKS (existence, differentiability and the envelope condition, under their Assumption~C) as given; its proof does not use their Lemma~5. We add a regularity condition on the set of securities that DKS do not state. Lemma~5 of DKS is also invoked in the proofs of other results of that paper, on the comparison of auction formats and on informal auctions; these hold under the cross-partial condition, and we do not examine whether they hold under the likelihood ratio property alone. For \citet{Yuan2026} we take the equilibrium characterization of that paper as given, and add a regularity condition on the set of securities for the revenue comparison.

\section{Definitions}

Let $A\subseteq\R$ and $B\subseteq\R$ be intervals with nonempty interior, and $g:A\times B\to(0,\infty)$ continuous. For a bounded measurable $f$ with compact support in $A$, write $h(y)=\int_A f(x)g(x,y)\,dx$.

\begin{definition}
A function $\phi$ on an interval satisfies \emph{weak single crossing} (WSC) if $\phi(x)>0$ implies $\phi(x')\ge0$ for $x'>x$, and \emph{strict single crossing} (SSC) if $\phi(x)\ge0$ implies $\phi(x')>0$ for $x'>x$.
\end{definition}

Let $\W$ be the class of bounded measurable functions $f:A\to\R$ with compact support in $A$ that satisfy WSC and are non-zero on a set of positive measure. For $f\in\W$ all the integrals below exist. For $f\in\W$ define the threshold
\[
x_f=\inf\{x\in A: f(x)>0\},
\]
with $x_f=\sup A$ if the set is empty. By WSC, $f\ge0$ on $(x_f,\sup A)$, and $f\le0$ on $(\inf A,x_f)$ by definition.

\begin{definition}
For $y_1<y_2$ in $B$, the \emph{likelihood ratio} is $R_{y_1y_2}(x)=g(x,y_2)/g(x,y_1)$. When $g$ is differentiable in $y$, the \emph{score} is $q(x,y)=\partial_y\log g(x,y)$.
\end{definition}

\section{Preservation of strict single crossing}

\begin{theorem}\label{thm:A}
The following are equivalent:
\begin{enumerate}
\item[(i)] for every $f\in\W$ and $y_1<y_2$ in $B$, $\int_Afg(\cdot,y_1)\ge0$ implies $\int_Afg(\cdot,y_2)>0$;
\item[(ii)] for every $y_1<y_2$ in $B$, $R_{y_1y_2}$ is strictly increasing on $A$.
\end{enumerate}
\end{theorem}

\begin{proof}
(ii)$\Rightarrow$(i). Fix $f\in\W$, $y_1<y_2$, and suppose $h(y_1)\ge0$. If $f\ge0$ almost everywhere, then $h(y_2)>0$ because $f\ne0$ on a set of positive measure and $g>0$. Otherwise $\{f<0\}$ has positive measure; if $\{f>0\}$ were null, then $h(y_1)<0$, a contradiction. Hence both sets have positive measure, and $x_f$ is an interior point of $A$. Writing $R=R_{y_1y_2}$,
\[
h(y_2)=\int_A f(x)\big[R(x)-R(x_f)\big]g(x,y_1)\,dx+R(x_f)\,h(y_1).
\]
The integrand is non-negative, and positive wherever $f(x)\ne0$ and $x\ne x_f$, a set of positive measure; and $R(x_f)h(y_1)\ge0$. Hence $h(y_2)>0$.

(i)$\Rightarrow$(ii). Suppose $R=R_{y_1y_2}$ is not strictly increasing. We construct disjoint compact intervals $I_1<I_2$ in $A$, of positive length, on which the $g(\cdot,y_1)$-weighted averages $M_j=\int_{I_j}Rg(\cdot,y_1)\big/\int_{I_j}g(\cdot,y_1)$ satisfy $M_1\ge M_2$. Two cases are possible. If there are $x_1<x_2$ in $A$ with $R(x_1)>R(x_2)$, then, since $R$ is continuous and $A$ has nonempty interior, there are compact intervals $I_1\ni x_1$ and $I_2\ni x_2$ of positive length, contained in $A$, with $I_1<I_2$ and $\max_{I_1}\!\big(R(x_1)-R\big)$, $\max_{I_2}\!\big(R-R(x_2)\big)$ both smaller than $\tfrac13\Delta$, where $\Delta=R(x_1)-R(x_2)$; then $M_1>R(x_1)-\tfrac13\Delta=R(x_2)+\tfrac23\Delta>R(x_2)+\tfrac13\Delta>M_2$. (If $x_1$ is the left endpoint of $A$, take $I_1=[x_1,x_1+\varepsilon]$; similarly for $x_2$.) If there is no such pair, $R$ is non-decreasing on $A$, and, not being strictly increasing, it takes the same value at two points $x_1<x_2$, hence is constant on $[x_1,x_2]$; any disjoint compact intervals $I_1<I_2$ of positive length in $[x_1,x_2]$ give $M_1=M_2$. With $G_j=\int_{I_j}g(\cdot,y_1)>0$, let $f=-\mathbf 1_{I_1}+(G_1/G_2)\mathbf 1_{I_2}\in\W$. It is bounded, has compact support in $A$, satisfies WSC (it is positive only on $I_2$ and vanishes to the right of $I_2$), and is non-zero on a set of positive measure. Then $h(y_1)=0$ and $h(y_2)=G_1(M_2-M_1)\le0$, contradicting (i).
\end{proof}

Condition (ii) is the ratio form of strict log-supermodularity: $g(x_1,y_1)g(x_2,y_2)>g(x_1,y_2)g(x_2,y_1)$ for $x_1<x_2$ and $y_1<y_2$. It involves no derivatives.

\begin{remark}[Integrable $f$]\label{rem:integrable}
In the proof of (ii)$\Rightarrow$(i), boundedness and compact support of $f$ are used only to guarantee that the integrals exist. The implication therefore holds for every measurable $f$ that satisfies WSC, is non-zero on a set of positive measure, and is such that $\int_A|f|\,g(\cdot,y)<\infty$ for every $y$. The same applies to (ii)$\Rightarrow$(i) of Theorem~\ref{thm:B} below, when differentiation under the integral sign is permitted.
\end{remark}

\section{Transversality}

\begin{theorem}\label{thm:B}
Fix $y\in B$. Suppose $g$ is continuously differentiable in $y$, $q(\cdot,y)$ is continuous on $A$, and differentiation under the integral sign is permitted for $f\in\W$. The following are equivalent:
\begin{enumerate}
\item[(i)] for every $f\in\W$, $h(y)=0$ implies $h'(y)>0$;
\item[(ii)] $x\mapsto q(x,y)$ is strictly increasing on $A$.
\end{enumerate}
\end{theorem}

\begin{proof}
(ii)$\Rightarrow$(i). If $h(y)=0$ then, as in the proof of Theorem~\ref{thm:A}, $x_f$ is interior, and
\[
h'(y)=\int_A f\,g\,q\,dx=\int_A f(x)\big[q(x,y)-q(x_f,y)\big]g(x,y)\,dx>0 .
\]
(i)$\Rightarrow$(ii). If $q(\cdot,y)$ is not strictly increasing, the case distinction in the proof of Theorem~\ref{thm:A}, applied to the continuous function $q(\cdot,y)$ in place of $R$, gives disjoint compact intervals $I_1<I_2$ of positive length in $A$ on which the $g(\cdot,y)$-weighted averages $M_j$ of $q(\cdot,y)$ satisfy $M_1\ge M_2$. For $f=-\mathbf 1_{I_1}+(G_1/G_2)\mathbf 1_{I_2}$, $h(y)=0$ and $h'(y)=G_1(M_2-M_1)\le0$.
\end{proof}

\begin{proposition}\label{prop:hier}
Under the regularity of Theorem~\ref{thm:B} for all $y$:
\begin{enumerate}
\item[(a)] if $q(\cdot,y)$ is strictly increasing for every $y$, then every $R_{y_1y_2}$ is strictly increasing;
\item[(b)] every $R_{y_1y_2}$ strictly increasing implies that $q(\cdot,y)$ is non-decreasing for every $y$, but not that it is strictly increasing.
\end{enumerate}
\end{proposition}

\begin{proof}
(a) $\log R_{y_1y_2}(x)=\int_{y_1}^{y_2}q(x,t)\,dt$, an integral of strictly increasing functions of $x$. (b) For $x_1<x_2$, $t\mapsto\log R_{y,t}(x_2)-\log R_{y,t}(x_1)$ is positive for $t>y$ and vanishes at $t=y$, so its right derivative at $t=y$, which is $q(x_2,y)-q(x_1,y)$, is non-negative. If $y=\sup B$, apply the same argument to $t\mapsto\log R_{t,y}(x_2)-\log R_{t,y}(x_1)$ for $t<y$, which is positive and vanishes at $t=y$, so that its left derivative at $t=y$, which is $-\big(q(x_2,y)-q(x_1,y)\big)$, is non-positive. Example~\ref{ex:sep} shows that it can vanish.
\end{proof}

\begin{corollary}[Weak ranking]\label{cor:weak}
Under the regularity of Theorem~\ref{thm:B} for all $y$, suppose every $R_{y_1y_2}$ is strictly increasing. Then for every $f\in\W$, $h(y)=0$ implies $h'(y)\ge0$.
\end{corollary}

\begin{proof}
By Proposition~\ref{prop:hier}(b), $q(\cdot,y)$ is non-decreasing, so the integrand in the expression for $h'(y)$ in the proof of Theorem~\ref{thm:B} is non-negative.
\end{proof}

\begin{example}\label{ex:sep}
Let $A=[0,2]$, $g(x,y)=e^{xy^3}$ and $f=-\mathbf 1_{[0,1]}+\mathbf 1_{(1,2]}\in\W$. For $x_1<x_2$ and $y_1<y_2$, $\log\big[g(x_1,y_1)g(x_2,y_2)/g(x_1,y_2)g(x_2,y_1)\big]=(x_2-x_1)(y_2^3-y_1^3)>0$, so condition (ii) of Theorem~\ref{thm:A} holds. But $q(x,y)=3xy^2$ vanishes identically at $y=0$. Here $h(0)=0$ and $h'(0)=0$, while $h(y)=y^3+o(y^3)$ near $0$, so $h$ changes sign from negative to positive at $0$ with zero derivative.
\end{example}

\begin{remark}[The phenomenon at the level of primitives]\label{rem:exp}
Let $Z\,|\,V=v$ be exponential with mean $\theta(v)=c+(v-m)^3$, where $m$ is interior to the support $[v_L,v_H]$ of $V$ and $c>(m-v_L)^3$. Since $\theta$ is strictly increasing, the likelihood ratio $f_{Z|V}(z|v)/f_{Z|V}(z|v')$ is strictly increasing in $z$ for $v>v'$: the pair $(Z,V)$ satisfies the strict monotone likelihood ratio property in ratio form. But $\partial_v\log f_{Z|V}(z|v)=\theta'(v)\big(z/\theta(v)^2-1/\theta(v)\big)$ vanishes for every $z$ at $v=m$. Hence for every payoff difference $G$ with $E[G(Z)|V=m]=0$, the derivative of $v\mapsto E[G(Z)|V=v]$ at $m$ is zero; if $G$ is weakly single crossing and non-zero on a set of positive measure, Theorem~\ref{thm:A} shows that the sign of $E[G(Z)|V=v]$ nevertheless changes there, from negative to positive.
\end{remark}

\section{Crossing indices and a local-to-global principle}\label{sec:index}

The previous section shows that transversality is strictly stronger than what likelihood ratios deliver. What they do deliver can be described locally, without derivatives.

\begin{definition}
Let $h$ be continuous on an interval $B$ and $h(y_0)=0$ with $y_0$ interior. The zero $y_0$ has \emph{index} $+1$ if there is $\delta>0$ with $h<0$ on $(y_0-\delta,y_0)$ and $h>0$ on $(y_0,y_0+\delta)$, and index $-1$ if the signs are reversed.
\end{definition}

A transversal zero, $h'(y_0)>0$, has index $+1$, but so does the zero of $y^3$ at the origin.

\begin{theorem}\label{thm:index}
Let $h$ be continuous on an open interval $B$. Then $h$ satisfies SSC if and only if $h$ has at most one zero and, if it has one, that zero has index $+1$. Moreover, if every zero of $h$ is isolated and has index $+1$, then $h$ has at most one zero, and hence satisfies SSC.
\end{theorem}

\begin{proof}
Suppose $h$ satisfies SSC and $h(y_0)=0$. Then $h>0$ on $(y_0,\sup B)$, and $h<0$ on $(\inf B,y_0)$, because $h(y)\ge0$ for some $y<y_0$ would force $h(y_0)>0$. So $y_0$ is the only zero and has index $+1$. Conversely, if $h$ has no zero it has constant sign and SSC holds; if it has a single zero of index $+1$, the intermediate value theorem gives $h<0$ to its left and $h>0$ to its right, and SSC holds.

For the second statement, suppose every zero is isolated with index $+1$ and that $y_1<y_2$ are zeros. The zero set in $[y_1,y_2]$ is compact and discrete, hence finite, so we may take $y_1,y_2$ consecutive. On $(y_1,y_2)$, $h$ has no zero and therefore constant sign. Index $+1$ at $y_1$ makes it positive, and index $+1$ at $y_2$ makes it negative, a contradiction.
\end{proof}

The proof of Theorem~\ref{thm:index} is elementary: it uses only continuity, the isolation of the zeros and the intermediate value theorem. It can be read as the one-dimensional case of a degree argument, in which the sum of the local indices of the zeros is determined by the behaviour of $h$ at the boundary of $B$, so that if every index equals $+1$ there is at most one zero. In general equilibrium this is the index theorem of \citet{Dierker1972}, and the uniqueness criterion of \citet{Varian1975}: if every equilibrium of a regular economy has index $+1$, the equilibrium is unique. There the index is the sign of a Jacobian determinant and requires regularity, that is, transversality; here it is defined by the change of sign, and a zero of index $+1$ may have zero derivative. A degree-theoretic extension to several parameters is left for future work.

For $f\in\W$, the function $h$ is continuous on $B$: if $K\subset A$ is the compact support of $f$ and $y_n\to y$ in $B$, then $g$ is bounded on $K\times J$ for a compact neighbourhood $J$ of $y$ in $B$, and dominated convergence applies. Combined with Theorem~\ref{thm:A}:

\begin{corollary}\label{cor:index}
Let $B$ be open. If every likelihood ratio of $g$ is strictly increasing, then for every $f\in\W$ the function $h$ has at most one zero, and every zero of $h$ has index $+1$. Transversality at the zero holds for every $f\in\W$ if and only if, in addition, the condition of Theorem~\ref{thm:B} holds there.
\end{corollary}

The practical lesson is that arguments which only need the global sign pattern, such as ``the switch happens once and from below'', should be run with indices, which likelihood ratios deliver, rather than with derivatives, which require the score condition. Arguments that genuinely need a positive derivative at the switch, such as perturbation and implicit-function arguments, require Theorem~\ref{thm:B}.

\section{Security-bid auctions}\label{sec:dks}

\subsection{The model of DKS}

We use the model and, except for the conditional density, the notation of \citet{DeMarzoKremerSkrzypacz2005}. There are $n$ bidders competing for a project that requires an investment $X$. Bidder $i$ has a private signal $V_i$; signals are i.i.d. with distribution $F$ and positive density on $[v_L,v_H]$. Conditional on $V_i=v$, the payoff $Z_i$ of the project has a density $k(z|v)$ (denoted $h(z|v)$ in DKS) with full support on $[0,\infty)$, twice differentiable in $z$ and $v$, and:
\begin{itemize}
\item[] \emph{SMLRP} (DKS, Assumption~A(c)): for $v>v'$, the likelihood ratio $k(z|v)/k(z|v')$ is strictly increasing in $z$.
\end{itemize}
The signal is normalized so that $E[Z_i|V_i=v]-X=v$. A \emph{feasible security} is a function $S(z)$ such that $S$ and $z-S(z)$ are non-decreasing and $0\le S(z)\le z$. An \emph{ordered set of securities} is a family $S(s,\cdot)$, $s\in[s_0,s_1]$, of feasible securities such that $ES(s,v)\equiv E[S(s,Z)|V=v]$ satisfies $\partial_sES>0$, $ES(s_0,v_L)\le v_L$ and $ES(s_1,v_H)\ge v_H$. A bidder of type $v$ who wins and pays the security $S(s,\cdot)$ obtains $v-ES(s,v)$.

DKS say that a security $S_1$ \emph{strictly crosses} $S_2$ \emph{from below} if $E[S_1(Z)|v^*]=E[S_2(Z)|v^*]$ implies $\frac{d}{dv}E[S_1(Z)|v]\big|_{v^*}>\frac{d}{dv}E[S_2(Z)|v]\big|_{v^*}$, and that an ordered set $\mathcal S_1$ is \emph{steeper} than $\mathcal S_2$ if every $S_1\in\mathcal S_1$ strictly crosses every $S_2\in\mathcal S_2$ from below. Their Lemma~5 states that a sufficient condition for $S_1$ to strictly cross $S_2$ from below is that $S_1\ne S_2$ and that there is $z^*$ with $S_1(z)\le S_2(z)$ for $z<z^*$ and $S_1(z)\ge S_2(z)$ for $z>z^*$. Their Proposition~1 states that if $\mathcal S_1$ is steeper than $\mathcal S_2$, then in a first-price or second-price auction, for any realization of types, the seller's revenue is higher with $\mathcal S_1$.

For a pair of securities as in Lemma~5 of DKS, $f=S_1-S_2$ satisfies WSC and is non-zero on a set of positive measure: since $S$ and $z-S(z)$ are non-decreasing, every feasible security is $1$-Lipschitz and hence continuous, so two distinct feasible securities differ on an open set; and the difference of expected payments is $D(v)=\int_0^\infty f(z)k(z|v)\,dz$, an integral of the form studied above, with $|f(z)|\le z$ integrable. The SMLRP is condition (ii) of Theorem~\ref{thm:A}. Footnote~4 of DKS notes that it is equivalent to log-supermodularity of the density, which can be written as $\partial^2\log k/\partial z\,\partial v>0$ under differentiability. With weak inequalities this equivalence is standard. With strict inequalities, a positive cross-partial derivative makes the score $\partial_v\log k(\cdot|v)$ strictly increasing, which is condition (ii) of Theorem~\ref{thm:B} and implies the SMLRP; the converse fails (Proposition~\ref{prop:hier}). DKS state that their assumptions allow derivatives to be taken through expectation operators; we maintain this as an assumption, namely that differentiation in $v$ under the integral sign is permitted for differences of feasible securities. The results of the previous sections therefore give:

\begin{corollary}\label{cor:dks}
\begin{enumerate}
\item[(a)] Under the SMLRP, for every pair of securities as in Lemma~5 of DKS, the difference $D$ of expected payments satisfies SSC, and $D(v^*)=0$ implies $D'(v^*)\ge0$.
\item[(b)] If, in addition, the score $\partial_v\log k(z|v)$ is strictly increasing in $z$ for every $v$, in particular if $\partial^2\log k/\partial z\,\partial v>0$, then $S_1$ strictly crosses $S_2$ from below: Lemma~5 of DKS holds.
\item[(c)] If at some $v$ the score is not strictly increasing in $z$, there is a bounded $f$ with compact support, satisfying WSC and non-zero on a set of positive measure, with $\int fk(\cdot|v)=0$ and $\frac{d}{dv}\int fk(\cdot|v)\le0$.
\end{enumerate}
\end{corollary}

\begin{proof}
(a) Theorem~\ref{thm:A}, Corollary~\ref{cor:weak} and Remark~\ref{rem:integrable}. (b) Theorem~\ref{thm:B} and Remark~\ref{rem:integrable}. (c) Theorem~\ref{thm:B}.
\end{proof}

The proof of Lemma~5 of DKS writes the derivative of $D$, at a point where $D$ vanishes, as the integral of $S_1-S_2$ times the difference between the score at $z$ and at the crossing point, and asserts that, by the SMLRP, the score is strictly increasing in $z$, so that the integrand is strictly positive wherever the securities differ.\footnote{~The proofs of the technical lemmas of the published version are in the AER Web Appendix. In the working-paper version \citep{DeMarzoKremerSkrzypacz2004}, where the proof of Lemma~5 is given, the step reads ``From SMLRP, $[h_v/h]$ is increasing in $z$''; by footnote~3 of DKS, ``increasing'' is meant in the strict sense.} This is the argument of Theorem~\ref{thm:B}: the strictness requires the score to be strictly increasing, which the cross-partial condition implies, and by Corollary~\ref{cor:dks}(c) nothing weaker than strict monotonicity of the score suffices.

\subsection{The SMLRP does not imply the ranking of slopes}

Part (c) of Corollary~\ref{cor:dks} does not by itself show that Lemma~5 of DKS can fail under their assumptions, because the function $f$ must be a difference of feasible securities and the density must satisfy the normalization of the signal. Under that normalization the score $\partial_v\log k(\cdot|v)$ cannot vanish identically at any $v$, since its covariance with $Z$ is $1$, so a kernel like that of Example~\ref{ex:sep} cannot be used. A score that is constant on an interval suffices.

\begin{example}[Inside the model of DKS]\label{ex:dks}
Let $\gamma(s)=s^3/(1+s^3)$ for $s>0$ and $\gamma(s)=0$ for $s\le0$, a bounded $C^2$ function, and put
\[
\phi(z)=\gamma(z-3)-\gamma(1-z),\qquad \psi(z)=\arctan z .
\]
Then $\phi$ is bounded, $C^2$ and non-decreasing, with $\phi'=0$ exactly on $[1,3]$ and $\phi\equiv0$ there, and $\psi$ is bounded with $\psi'>0$. For $t\in[-1,1]$ let
\[
k_0(z|t)=\frac{1}{N(t)}\exp\!\big(-z+t\,\phi(z)+t^3\psi(z)\big),\qquad z\ge0,
\]
where $N(t)$ normalizes; $N(0)=1$. The density has full support on $[0,\infty)$, is $C^2$ in $z$ and $C^\infty$ in $t$.
\begin{enumerate}
\item[(a)] \emph{Strict MLRP.} For $t_1<t_2$, $\partial_z\log\big[k_0(z|t_2)/k_0(z|t_1)\big]=(t_2-t_1)\phi'(z)+(t_2^3-t_1^3)\psi'(z)>0$ for every $z$.
\item[(b)] \emph{Normalization.} Let $m(t)=E[Z|t]$. Then $m'(t)=\mathrm{Cov}_t\big(Z,\phi(Z)+3t^2\psi(Z)\big)>0$, as the covariance of two non-decreasing, non-constant functions of $Z$ under a density with full support. Hence $t\mapsto v=m(t)-X$, with $X\le m(-1)$, is a $C^\infty$ diffeomorphism onto an interval $[v_L,v_H]\subset[0,\infty)$, and $k(z|v)=k_0(z|t(v))$ satisfies all the assumptions above, with $E[Z|V=v]-X=v$.
\item[(c)] \emph{Securities.} Let $\beta(s)=(1-4s^2)^2$ for $|s|\le\tfrac12$ and $\beta(s)=0$ otherwise, so that $|\beta'|\le 16/(3\sqrt3)$. Put $S_2(z)=z/2$ and $S_1(z)=z/2+\delta(z)$ with
\[
\delta(z)=-\tfrac1{20}\,\beta\big(z-\tfrac32\big)+\tfrac{e}{20}\,\beta\big(z-\tfrac52\big).
\]
Since $|\delta'|<\tfrac12$ and $|\delta|<\tfrac12\le z/2$ on the support $[1,3]$ of $\delta$, both $S_j$ and $z-S_j$ are non-decreasing and $0\le S_j\le z$: both are feasible securities. Moreover $S_1\ne S_2$, $S_1\le S_2$ for $z<2$ and $S_1\ge S_2$ for $z>2$, so the pair satisfies the hypothesis of Lemma~5 of DKS.
\item[(d)] \emph{Failure of strict crossing.} Let $D(t)=E[S_1(Z)-S_2(Z)\,|\,t]=\int\delta\,k_0(\cdot|t)$. At $t=0$, $k_0(z|0)=e^{-z}$, and $\int\beta(z-\tfrac52)e^{-z}dz=e^{-1}\int\beta(z-\tfrac32)e^{-z}dz$, so $D(0)=0$. Since $\partial_t\log k_0(z|0)=\phi(z)-E_0[\phi(Z)]$ is constant on $[1,3]$, $D'(0)=\big(0-E_0[\phi(Z)]\big)\int\delta\,e^{-z}dz=0$. In the normalized signal, $dD/dv=D'(t)/m'(t)=0$ at $v_0=m(0)-X$, although $E[S_1|v_0]=E[S_2|v_0]$.
\end{enumerate}
Thus $S_1$ does not strictly cross $S_2$ from below in the sense of DKS (equality of expected payoffs implying a strict ranking of their derivatives), although all their assumptions hold. Consistently with Theorem~\ref{thm:A}, the difference $D$ still satisfies strict single crossing: it changes sign once, from negative to positive, with zero derivative. More precisely, since $\phi\equiv0$ on the support $[1,3]$ of $\delta$ and $\int\delta\,e^{-z}dz=0$,
\[
D(t)=\frac{1}{N(t)}\int_1^3\delta(z)\,e^{-z}\big(e^{t^3\arctan z}-1\big)\,dz=c\,t^3+O(t^4),\qquad c=\int_1^3\delta(z)\,e^{-z}\arctan z\,dz .
\]
Substituting $z=u+1$ in the part of $\delta$ supported on $[2,3]$,
\[
c=\frac1{20}\int_1^2\beta\big(u-\tfrac32\big)\,e^{-u}\big[\arctan(u+1)-\arctan u\big]\,du>0,
\]
since $\beta\ge0$, $\beta>0$ on $(1,2)$ and $\arctan$ is strictly increasing. Hence $D(0)=D'(0)=D''(0)=0<D'''(0)$; numerically, $c\approx1.315\times10^{-3}$.
\end{example}

\subsection{The revenue ranking holds under the SMLRP}

We now show that the ranking of revenues in Proposition~1 of DKS does not depend on the strict ranking of slopes. Let $\mathcal S_1$ and $\mathcal S_2$ be ordered sets of securities, with expected payments $ES^1$ and $ES^2$, such that, for every $S_1\in\mathcal S_1$ and $S_2\in\mathcal S_2$,
\begin{equation}\label{eq:wscpairs}
S_1-S_2\ \text{satisfies WSC and }S_1\ne S_2\text{ on a set of positive measure}.
\end{equation}
For pairs that are not ordered pointwise this is the hypothesis of Lemma~5 of DKS: the payoffs cross once from below.

\begin{proposition}[Second-price auctions]\label{prop:spa}
Assume the SMLRP and \eqref{eq:wscpairs}. Fix $v_2$ and let $S_j\in\mathcal S_j$ be the security bid by type $v_2$ in the second-price auction with the set $\mathcal S_j$, so that $E[S_j(Z)|v_2]=v_2$, $j=1,2$. Then $E[S_1(Z)|v_1]>E[S_2(Z)|v_1]$ for every $v_1>v_2$: for every realization of types in which the winner's type exceeds the second highest, the seller's revenue is strictly higher with $\mathcal S_1$.
\end{proposition}

\begin{proof}
In a second-price auction a bidder of type $v$ bids the security whose expected payment at $v$ equals $v$ (DKS, Lemma~2), and the winner pays the security bid by the second highest type. With $f=S_1-S_2$ and $D(v)=\int fk(\cdot|v)$, $D(v_2)=v_2-v_2=0$, and Corollary~\ref{cor:dks}(a) gives $D(v_1)>0$ for $v_1>v_2$.
\end{proof}

The same observation applies to later work that adopts the derivative notion of steepness of DKS. In the second-price auctions with competing sellers of \citet{GorbenkoMalenko2011} and the auctions with endogenous entry of \citet{SogoBernhardtLiu2016}, the strict ranking of slopes at a point of equal expected payments is used to order expected payments above that point. By Theorem~\ref{thm:A}, this ordering holds under the SMLRP whenever the payoffs of the securities cross once from below, without the strict ranking of slopes.

For first-price auctions we take the characterization of equilibrium in DKS as given. Under their Assumption~C, with the set $\mathcal S_j$ there is a unique symmetric equilibrium bidding strategy $b_j(\cdot)$, which is increasing and differentiable, with $ES^j(b_j(v_L),v_L)=v_L$ (DKS, Lemma~3). Let $P^j(v)=ES^j(b_j(v),v)$ be the equilibrium expected payment of a winner of type $v$ and
\[
U^j(v)=F^{n-1}(v)\big(v-P^j(v)\big)
\]
the equilibrium payoff. Then $U^j$ is differentiable, $U^j(v_L)=0$, and the envelope condition holds (DKS, equation~(3)):
\begin{equation}\label{eq:env}
U^{j\prime}(v)=F^{n-1}(v)\big(1-\partial_vES^j(b_j(v),v)\big).
\end{equation}
For $p$ in the range of $ES^j(\cdot,v)$, let $s^j(p,v)$ be the unique $s$ with $ES^j(s,v)=p$, and define
\[
\sigma^j(v,p)=\partial_vES^j\big(s^j(p,v),v\big),
\]
the slope, with respect to the true type, of the expected payment of the security of $\mathcal S_j$ whose expected payment at $v$ is $p$. Then \eqref{eq:env} reads $U^{j\prime}(v)=F^{n-1}(v)\big(1-\sigma^j(v,P^j(v))\big)$.

\begin{lemma}[Weak ranking of slopes]\label{lem:weak}
Assume the SMLRP and \eqref{eq:wscpairs}. Then $\sigma^1(v,p)\ge\sigma^2(v,p)$ whenever both are defined. If the score $\partial_v\log k(\cdot|v)$ is strictly increasing at $v$, the inequality is strict.
\end{lemma}

\begin{proof}
The securities $S_j=S^j(s^j(p,v),\cdot)$ have the same expected payment $p$ at $v$, so $D(v)=0$ for $f=S_1-S_2$, and $\sigma^1(v,p)-\sigma^2(v,p)=D'(v)$. Apply Corollary~\ref{cor:dks}(a) and (b).
\end{proof}

\begin{proposition}[First-price auctions]\label{prop:fpa}
Assume the SMLRP, \eqref{eq:wscpairs}, and Assumption~C of DKS for both sets of securities, so that the first-price equilibria are characterized as above. Suppose that $ES^2$ is twice continuously differentiable on $[s_0,s_1]\times[v_L,v_H]$, with $\partial_sES^2>0$, and that $ES^2(s_0,v)\le v_L$ and $ES^2(s_1,v)\ge v$ for every $v$. Then, in the first-price auctions with the sets $\mathcal S_1$ and $\mathcal S_2$:
\begin{enumerate}
\item[(a)] $U^1(v)\le U^2(v)$ for every $v\in[v_L,v_H]$: for each realization of the winning type, the seller's expected revenue is at least as high with $\mathcal S_1$ as with $\mathcal S_2$;
\item[(b)] if the score $\partial_v\log k(\cdot|v)$ is strictly increasing in $z$ for almost every $v$, then $U^1(v)<U^2(v)$ for every $v>v_L$.
\end{enumerate}
\end{proposition}

\begin{proof}
\emph{Step 1: payments.} Since $U^1\ge0$, $P^1(v)\le v$. Since $b_1$ is increasing, $ES^1$ is increasing in $s$, and $ES^1(s,\cdot)$ is non-decreasing (the security is non-decreasing and the SMLRP holds), $P^1(v)\ge ES^1(b_1(v_L),v)\ge ES^1(b_1(v_L),v_L)=v_L$. Hence $P^1(v)\in[v_L,v]$ lies in the range of $ES^2(\cdot,v)$, and so does $P^2(v)$; $\sigma^2(v,p)$ is defined for $p$ between them.

\emph{Step 2: a Lipschitz bound.} For fixed $v$, $\partial_p\sigma^2(v,p)=\partial^2_{sv}ES^2/\partial_sES^2$, evaluated at $(s^2(p,v),v)$. By continuity on the compact set $[s_0,s_1]\times[v_L,v_H]$, there is $L$ with $|\sigma^2(v,p)-\sigma^2(v,p')|\le L|p-p'|$ for all $v$ and all $p,p'$ in the range of $ES^2(\cdot,v)$.

\emph{Step 3: a differential inequality.} Let $G=F^{n-1}$, $\Delta=U^2-U^1$ and $\omega(v)=\sigma^1(v,P^1(v))-\sigma^2(v,P^1(v))$, which is non-negative by Lemma~\ref{lem:weak}. Since $U^j=G\,(v-P^j)$, we have $\Delta=G\,(P^1-P^2)$. By \eqref{eq:env} and Step~2,
\begin{align*}
\Delta'&=G\big[\sigma^1(v,P^1)-\sigma^2(v,P^2)\big]=G\,\omega+G\big[\sigma^2(v,P^1)-\sigma^2(v,P^2)\big]\\
&\ge G\,\omega-L\,G\,|P^1-P^2|=G\,\omega-L\,|\Delta| .
\end{align*}

\emph{Step 4: comparison.} Suppose $\Delta(\hat v)<0$ for some $\hat v$, and let $v_0=\sup\{v<\hat v:\Delta(v)\ge0\}$, which exists because $\Delta(v_L)=0$. Then $\Delta(v_0)=0$ and $\Delta<0$ on $(v_0,\hat v]$, where $\Delta'\ge L\Delta$, that is, $\big(e^{-Lv}\Delta\big)'\ge0$. Hence $e^{-L\hat v}\Delta(\hat v)\ge e^{-Lv_0}\Delta(v_0)=0$, a contradiction. This proves (a). Then $\Delta'\ge G\,\omega-L\Delta$, so that
\[
\Delta(v)\ \ge\ e^{-Lv}\int_{v_L}^{v}e^{Ls}\,G(s)\,\omega(s)\,ds .
\]
Under the hypothesis of (b), $\omega>0$ almost everywhere by Lemma~\ref{lem:weak}, and $G>0$ on $(v_L,v_H]$, so $\Delta(v)>0$ for $v>v_L$.
\end{proof}

\begin{remark}[Pointwise and comparison arguments]
The proof of Proposition~1 in DKS argues pointwise: at a type where $U^1=U^2$, the expected payments coincide, the strict ranking of slopes gives $U^{1\prime}<U^{2\prime}$, and therefore $U^1<U^2$ above $v_L$. Without strictness this argument does not close, because a function may vanish with zero derivative and then change sign. The comparison argument uses the weak ranking of slopes at every payment level, not only where the payoffs coincide; the Lipschitz bound of Step~2 controls the slopes where the payments differ. The assumptions on $ES^2$ in Proposition~\ref{prop:fpa} are conditions on the set of securities and the density; they may equally be imposed on $ES^1$, exchanging the roles of the two sets in Step~3.
\end{remark}

\begin{remark}[Strictness]
In Example~\ref{ex:dks} the score fails to be strictly increasing at a single signal, so part (b) of Proposition~\ref{prop:fpa} applies: Lemma~5 of DKS fails there, yet the revenue ranking is strict. More generally, the ranking can fail to be strict only if the slopes coincide along the equilibrium path on a set of types of positive measure.
\end{remark}

To summarize: under Assumption~A(c) of DKS as stated, the derivative-based notion of steepness does not follow from the crossing of payoffs, but the ranking of revenues of their Proposition~1 does, for second-price auctions without further assumptions and for first-price auctions under the regularity condition of Proposition~\ref{prop:fpa}; under the cross-partial condition of their footnote~4, Lemma~5 and all the results that use it hold as stated.

\subsection{\texorpdfstring{Common values: the lemma in \citet{Yuan2026}}{Common values: the lemma in Yuan (2026)}}\label{sec:yuan-use}

Lemma~5 of \citet{Yuan2026} extends the lemma of DKS to the common-value setting of that paper, and its proof follows theirs. It states that if $f$ satisfies WSC and is non-zero on a set of positive measure, and $g>0$ is strictly log-supermodular, then $h$ satisfies SSC and $h(y)=0$ implies $h'(y)>0$, with strict log-supermodularity defined in ratio form. By Theorem~\ref{thm:A}, the first conclusion holds, and the hypothesis is the weakest possible for it to hold for every $f\in\W$. The second conclusion is the one inherited from DKS; by Theorem~\ref{thm:B} it requires the score condition, and Remark~\ref{rem:exp} shows that the score condition is not implied by the strict monotone likelihood ratio property for primitive densities satisfying Assumption~1 of \citet{Yuan2026}, which imposes no normalization of the signal. Corollary~\ref{cor:index} gives what the ratio form does deliver at a zero of $h$: a change of sign from negative to positive.

The first conclusion is what the main results of \citet{Yuan2026} use: the incentive compatibility of candidate monotonic strategies and the arguments behind Corollary~3(i), (ii) and (iv) of that paper, which drive the equilibrium characterizations, rely on SSC only (Online Appendix, Section~1). The second conclusion is used in Corollary~3(iii), a statement about the derivative of a conditional expectation at a point where it vanishes, and Corollary~3(iii) is used in the proof of Corollary~1 (Online Appendix, Section~4). That corollary states that, under the hypotheses of the paper, the entrepreneur's revenue in the informal auction is not lower than in any formal first-price auction restricted to an ordered set; the proof shows that, among formal first-price auctions, revenue is highest when offers are restricted to debts.

We recall the structure of that proof. Let $\mathcal S^D$ be the ordered set of debts and $\mathcal S^A=\{S^A(s,\cdot):s\in[s_L,s_H]\}$ another ordered set, with candidate monotonic strategies $s^D_x$ and $s^A_x$. Let $\Pi^D(x)$ and $\Pi^A(x)$ be the equilibrium payoffs of a bidder of type $x$, $F(x)=F_{Y|X}(x|x)$, and
\begin{align*}
\alpha(s,s',x)&=E\big[S^A(s,Z)-S^D(s',Z)\,\big|\,X_1=x,\ Y_1\le x\big],\\
\beta(s,s',x)&=\frac{\partial}{\partial x}E\big[S^A(s,Z)-S^D(s',Z)\,\big|\,X_1=x,\ Y_1\le\tilde x\big]\Big|_{\tilde x=x}.
\end{align*}
In \citet{Yuan2026}, $\Pi^A(x)-\Pi^D(x)=\alpha(s^A_x,s^D_x,x)\,F(x)$, this difference vanishes at $x_L$, and, by the envelope theorem,
\begin{equation}\label{eq:yuanenv}
\frac{d}{dx}\big(\Pi^A-\Pi^D\big)(x)=\beta(s^A_x,s^D_x,x)\,F(x)+\alpha(s^A_x,s^D_x,x)\,\frac{\partial F_{Y|X}(\tilde x|x)}{\partial x}\Big|_{\tilde x=x},
\end{equation}
where the last partial derivative is non-positive. The proof then needs $\beta\ge0$ whenever $\alpha$ lies in some interval $(-\epsilon,0)$. It obtains this from Corollary~3(iii), which gives $\beta>0$ where $\alpha=0$, and from continuity, choosing $\epsilon_x>0$ for each $x$ and setting $\epsilon=\inf_x\epsilon_x$; the conclusion then follows from a generalization of Lemma~2 of \citet{MilgromWeber1982}. This route goes through the strict inequality of Corollary~3(iii), which by Theorem~\ref{thm:B} rests on the score condition. The comparison argument of Proposition~\ref{prop:fpa} provides an alternative route, which uses only the weak ranking of slopes and a uniform bound on the set of securities. Write $m^A(s,x,\tilde x)=E[S^A(s,Z)\,|\,X_1=x,\ Y_1\le\tilde x]$.

\begin{proposition}\label{prop:yuan}
Assume Assumption~1 of \citet{Yuan2026}, that all ordered sets satisfy Condition~1 of that paper, and that one of the two cases of its Corollary~1 holds: either the entrepreneur observes $V$ and $\gamma^D(V)$ is connected for each $V$, or the entrepreneur has no private information and, if $N\ge3$, Assumption~2 of that paper holds and $\gamma_x(g)$ is connected for each $(x,g)$. These are the hypotheses of Corollary~1 of \citet{Yuan2026}; Corollary~3(iii) of that paper is not used. Suppose, in addition, that the conditional density of $Z$ given $(X_1=x,\ Y_1\le\tilde x)$ is differentiable in $x$, with differentiation under the integral sign permitted for differences of feasible securities, and that
\begin{enumerate}
\item[(R1)] $m^A$ is twice continuously differentiable on $[s_L,s_H]\times[x_L,x_H]^2$, and $\partial_sm^A\ge\kappa>0$ there;
\item[(R2)] for every $x$, the range of $s\mapsto m^A(s,x,x)$ contains $E\big[S^D(s^D_x,Z)\,\big|\,X_1=x,\ Y_1\le x\big]$, the expected payment of the debt bid by type $x$.
\end{enumerate}
Then $\Pi^A(x)\ge\Pi^D(x)$ for every $x\in[x_L,x_H]$, and the entrepreneur's revenue in the formal first-price auction restricted to $\mathcal S^A$ is not higher than in the one restricted to debts.
\end{proposition}

\begin{proof}
\emph{Step 1: a zero of $\alpha$ and the weak ranking there.} Fix $x$ and write $s'=s^D_x$. By (R2), and since $m^A$ is strictly increasing in $s$ by (R1), there is a unique $\hat s(x)\in[s_L,s_H]$ with $\alpha(\hat s(x),s',x)=0$. The securities $S^A(\hat s(x),\cdot)$ and $S^D(s',\cdot)$ are either ordered pointwise, in which case, since their expected payments coincide and the conditional density has full support, their difference vanishes almost everywhere and $\beta(\hat s(x),s',x)=0$, or not ordered, in which case the first crosses the debt from below and their difference satisfies WSC. In the second case, the pair $(Z,X_1)$ conditional on $Y_1\le x$ satisfies the SMLRP \citep[Lemma~3]{Yuan2026}, and Corollary~\ref{cor:weak}, with Remark~\ref{rem:integrable}, gives $\beta(\hat s(x),s',x)\ge0$.

\emph{Step 2: a Lipschitz bound.} Since the debt term of $\beta$ does not depend on $s$, $\partial_s\beta(s,s',x)=\partial^2_{s x}m^A(s,x,\tilde x)\big|_{\tilde x=x}$, which by (R1) is continuous, hence bounded, on the compact set $[s_L,s_H]\times[x_L,x_H]^2$. Thus $\beta(\cdot,s',x)$ is Lipschitz in $s$ with a constant $L_\beta$ independent of $(s',x)$, and $|s-\hat s(x)|\le \kappa^{-1}|\alpha(s,s',x)-\alpha(\hat s(x),s',x)|=\kappa^{-1}|\alpha(s,s',x)|$. Hence, with $L=L_\beta/\kappa$,
\[
\beta(s^A_x,s^D_x,x)\ \ge\ \beta(\hat s(x),s^D_x,x)-L\,|\alpha(s^A_x,s^D_x,x)|\ \ge\ -L\,|\alpha(s^A_x,s^D_x,x)| .
\]

\emph{Step 3: comparison.} Let $\Delta=\Pi^A-\Pi^D=\alpha F$, with $\alpha$ evaluated along the strategies. Where $\Delta<0$, $\alpha<0$ and $F>0$, so by \eqref{eq:yuanenv}, Step~2 and the sign of the last term in \eqref{eq:yuanenv},
\[
\Delta'\ \ge\ -L\,|\alpha|\,F=L\,\alpha F=L\,\Delta .
\]
Since $\Delta(x_L)=0$, the argument of Step~4 in the proof of Proposition~\ref{prop:fpa} gives $\Delta\ge0$ on $[x_L,x_H]$. Since the expected total surplus is the same in both formal auctions, the entrepreneur's revenue is not higher with $\mathcal S^A$.
\end{proof}

Proposition~\ref{prop:yuan} uses, in place of Corollary~3(iii), a uniform Lipschitz bound on the set of securities. Conditions (R1) and (R2) are on the ordered set and the distribution of $(Z,X_1,\dots,X_N)$; Since $0\le S^D(s^D_x,Z)\le Z$, (R2) holds whenever the range of $s\mapsto m^A(s,x,x)$ contains $\big[0,E[Z\,|\,X_1=x,\ Y_1\le x]\big]$, as for the ordered set of equity of \citet{Yuan2026}, $S^E(s,Z)=sZ$ with $s\in[0,1]$. For buyouts and call options it holds when the range of prices or strikes, starting at zero, is wide enough to reach the expected payment of the debt. The proof of Corollary~1 in \citet{Yuan2026} reduces the corollary to this ranking among formal first-price auctions; the rest of the argument rests on results of the main text, which, as noted, use only strict single crossing. Hence Corollary~1 of \citet{Yuan2026} holds under its hypotheses together with (R1), (R2) and the differentiability condition, imposed for every ordered set $\mathcal S^A$ with which debt is compared, without the score condition.

\section{Concluding remarks}

Likelihood ratios and scores carry different information. Strict monotonicity of likelihood ratios characterizes the preservation of strict single crossing, and therefore the global sign pattern; strict monotonicity of scores characterizes transversality at the crossing. The gap between them consists of crossings with zero derivative, which are invisible to derivative-based arguments but harmless for index-based ones. In security-bid auctions the distinction separates a lemma from the proposition it supports: the ranking of slopes needs the score condition, while the ranking of revenues survives under likelihood ratios, by single crossing for second-price auctions and by a comparison argument for first-price auctions. The same replacement of a pointwise transversality argument by a comparison argument may be useful in other applications of the linkage principle. Extending the index formulation to multidimensional, partially ordered parameter spaces through degree theory is a natural next step.

\end{document}